\documentclass[9pt]{article}
\usepackage{amsfonts,amssymb,amsmath,amscd,theorem,oldgerm,epsfig,units}
\usepackage[mathscr]{eucal}
\usepackage{mathrsfs}

\newcommand{\spc}[1]{\vskip-#1cm}

\makeatletter

\def\diagram{\m@th\leftwidth=\z@ \rightwidth=\z@ \topheight=\z@
\botheight=\z@ \setbox\@picbox\hbox\bgroup}

\def\enddiagram{\egroup\wd\@picbox\rightwidth\unitlength
\ht\@picbox\topheight\unitlength \dp\@picbox\botheight\unitlength
\hskip\leftwidth\unitlength\box\@picbox}

\def\bfig{\begin{diagram}}
\def\efig{\end{diagram}}
\newcount\wideness \newcount\leftwidth \newcount\rightwidth
\newcount\highness \newcount\topheight \newcount\botheight

\def\ratchet#1#2{\ifnum#1<#2 \global #1=#2 \fi}

\def\putbox(#1,#2)#3{%
\horsize{\wideness}{#3} \divide\wideness by 2
{\advance\wideness by #1 \ratchet{\rightwidth}{\wideness}}
{\advance\wideness by -#1 \ratchet{\leftwidth}{\wideness}}
\vertsize{\highness}{#3} \divide\highness by 2
{\advance\highness by #2 \ratchet{\topheight}{\highness}}
{\advance\highness by -#2 \ratchet{\botheight}{\highness}}
\put(#1,#2){\makebox(0,0){$#3$}}}

\def\putlbox(#1,#2)#3{%
\horsize{\wideness}{#3}
{\advance\wideness by #1 \ratchet{\rightwidth}{\wideness}}
{\ratchet{\leftwidth}{-#1}}
\vertsize{\highness}{#3} \divide\highness by 2
{\advance\highness by #2 \ratchet{\topheight}{\highness}}
{\advance\highness by -#2 \ratchet{\botheight}{\highness}}
\put(#1,#2){\makebox(0,0)[l]{$#3$}}}

\def\putrbox(#1,#2)#3{%
\horsize{\wideness}{#3}
{\ratchet{\rightwidth}{#1}}
{\advance\wideness by -#1 \ratchet{\leftwidth}{\wideness}}
\vertsize{\highness}{#3} \divide\highness by 2
{\advance\highness by #2 \ratchet{\topheight}{\highness}}
{\advance\highness by -#2 \ratchet{\botheight}{\highness}}
\put(#1,#2){\makebox(0,0)[r]{$#3$}}}

\def\adjust[#1]{} % For compatibility

\newcount \coefa
\newcount \coefb
\newcount \coefc
\newcount\tempcounta
\newcount\tempcountb
\newcount\tempcountc
\newcount\tempcountd
\newcount\xext
\newcount\yext
\newcount\xoff
\newcount\yoff
\newcount\gap%
\newcount\arrowtypea
\newcount\arrowtypeb
\newcount\arrowtypec
\newcount\arrowtyped
\newcount\arrowtypee
\newcount\height
\newcount\width
\newcount\xpos
\newcount\ypos
\newcount\run
\newcount\rise
\newcount\arrowlength
\newcount\halflength
\newcount\arrowtype
\newdimen\tempdimen
\newdimen\xlen
\newdimen\ylen
\newsavebox{\tempboxa}%
\newsavebox{\tempboxb}%
\newsavebox{\tempboxc}%

\newdimen\w@dth

\def\setw@dth#1#2{\setbox\z@\hbox{\m@th$#1$}\w@dth=\wd\z@
\setbox\@ne\hbox{\m@th$#2$}\ifnum\w@dth<\wd\@ne \w@dth=\wd\@ne \fi
\advance\w@dth by 1.2em}

\def\t@^#1_#2{\allowbreak\def\n@one{#1}\def\n@two{#2}\mathrel
{\setw@dth{#1}{#2}
\mathop{\hbox to \w@dth{\rightarrowfill}}\limits
\ifx\n@one\empty\else ^{\box\z@}\fi
\ifx\n@two\empty\else _{\box\@ne}\fi}}
\def\t@@^#1{\@ifnextchar_{\t@^{#1}}{\t@^{#1}_{}}}
\def\to{\@ifnextchar^{\t@@}{\t@@^{}}}

\def\t@left^#1_#2{\def\n@one{#1}\def\n@two{#2}\mathrel{\setw@dth{#1}{#2}
\mathop{\hbox to \w@dth{\leftarrowfill}}\limits
\ifx\n@one\empty\else ^{\box\z@}\fi
\ifx\n@two\empty\else _{\box\@ne}\fi}}
\def\t@@left^#1{\@ifnextchar_{\t@left^{#1}}{\t@left^{#1}_{}}}
\def\toleft{\@ifnextchar^{\t@@left}{\t@@left^{}}}

\def\two@^#1_#2{\allowbreak
\def\n@one{#1}\def\n@two{#2}\mathrel{\setw@dth{#1}{#2}
\mathop{\vcenter{\lineskip\z@\baselineskip\z@
                 \hbox to \w@dth{\rightarrowfill}%
                 \hbox to \w@dth{\rightarrowfill}}%
       }\limits
\ifx\n@one\empty\else ^{\box\z@}\fi
\ifx\n@two\empty\else _{\box\@ne}\fi}}
\def\tw@@^#1{\@ifnextchar _{\two@^{#1}}{\two@^{#1}_{}}}
\def\two{\@ifnextchar ^{\tw@@}{\tw@@^{}}}

\def\tofr@^#1_#2{\def\n@one{#1}\def\n@two{#2}\mathrel{\setw@dth{#1}{#2}
\mathop{\vcenter{\hbox to \w@dth{\rightarrowfill}\kern-1.7ex
                 \hbox to \w@dth{\leftarrowfill}}%
       }\limits
\ifx\n@one\empty\else ^{\box\z@}\fi
\ifx\n@two\empty\else _{\box\@ne}\fi}}
\def\t@fr@^#1{\@ifnextchar_ {\tofr@^{#1}}{\tofr@^{#1}_{}}}
\def\tofro{\@ifnextchar^ {\t@fr@}{\t@fr@^{}}}

\def\mon{\mathop{\m@th\hbox to
      14.6\P@{\lasyb\char'51\hskip-2.1\P@$\arrext$\hss
$\mathord\rightarrow$}}\limits} % width of \epi
\def\leftmono{\mathrel{\m@th\hbox to
14.6\P@{$\mathord\leftarrow$\hss$\arrext$\hskip-2.1\P@\lasyb\char'50%
}}\limits} % width of \epi
\mathchardef\arrext="0200       % amr minus for arrow extension (see \into)

\def\settypes(#1,#2,#3){\arrowtypea#1 \arrowtypeb#2 \arrowtypec#3}
\def\settoheight#1#2{\setbox\@tempboxa\hbox{#2}#1\ht\@tempboxa\relax}%
\def\settodepth#1#2{\setbox\@tempboxa\hbox{#2}#1\dp\@tempboxa\relax}%
\def\settokens`#1`#2`#3`#4`{%
     \def\tokena{#1}\def\tokenb{#2}\def\tokenc{#3}\def\tokend{#4}}
\def\setsqparms[#1`#2`#3`#4;#5`#6]{%
\arrowtypea #1
\arrowtypeb #2
\arrowtypec #3
\arrowtyped #4
\width #5
\height #6
}
\def\setpos(#1,#2){\xpos=#1 \ypos#2}

\def\settriparms[#1`#2`#3;#4]{\settripairparms[#1`#2`#3`1`1;#4]}%

\def\settripairparms[#1`#2`#3`#4`#5;#6]{%
\arrowtypea #1
\arrowtypeb #2
\arrowtypec #3
\arrowtyped #4
\arrowtypee #5
\width #6
\height #6
}

\def\resetparms{\settripairparms[1`1`1`1`1;500]\width 500}%default values%

\resetparms

\def\mvector(#1,#2)#3{%%
\put(0,0){\vector(#1,#2){#3}}%
\put(0,0){\vector(#1,#2){26}}%
}
\def\evector(#1,#2)#3{{%%
\arrowlength #3
\put(0,0){\vector(#1,#2){\arrowlength}}%
\advance \arrowlength by-30
\put(0,0){\vector(#1,#2){\arrowlength}}%
}}

\def\horsize#1#2{%
\settowidth{\tempdimen}{$#2$}%
#1=\tempdimen
\divide #1 by\unitlength
}

\def\vertsize#1#2{%
\settoheight{\tempdimen}{$#2$}%
#1=\tempdimen
\settodepth{\tempdimen}{$#2$}%
\advance #1 by\tempdimen
\divide #1 by\unitlength
}

\def\putvector(#1,#2)(#3,#4)#5#6{{%
\ifnum3<\arrowtype
\putdashvector(#1,#2)(#3,#4)#5\arrowtype
\else
\ifnum\arrowtype<-3
\putdashvector(#1,#2)(#3,#4)#5\arrowtype
\else
\xpos=#1
\ypos=#2
\run=#3
\rise=#4
\arrowlength=#5
\ifnum \arrowtype<0
    \ifnum \run=0
        \advance \ypos by-\arrowlength
    \else
        \tempcounta \arrowlength
        \multiply \tempcounta by\rise
        \divide \tempcounta by\run
        \ifnum\run>0
            \advance \xpos by\arrowlength
            \advance \ypos by\tempcounta
        \else
            \advance \xpos by-\arrowlength
            \advance \ypos by-\tempcounta
        \fi
    \fi
    \multiply \arrowtype by-1
    \multiply \rise by-1
    \multiply \run by-1
\fi
\ifcase \arrowtype
\or \put(\xpos,\ypos){\vector(\run,\rise){\arrowlength}}%
\or \put(\xpos,\ypos){\mvector(\run,\rise)\arrowlength}%
\or \put(\xpos,\ypos){\evector(\run,\rise){\arrowlength}}%
\fi\fi\fi
}}

\def\putsplitvector(#1,#2)#3#4{%%
\xpos #1
\ypos #2
\arrowtype #4
\halflength #3
\arrowlength #3
\gap 140
\advance \halflength by-\gap
\divide \halflength by2
\ifnum\arrowtype>0
   \ifcase \arrowtype
   \or \put(\xpos,\ypos){\line(0,-1){\halflength}}%
       \advance\ypos by-\halflength
       \advance\ypos by-\gap
       \put(\xpos,\ypos){\vector(0,-1){\halflength}}%
   \or \put(\xpos,\ypos){\line(0,-1)\halflength}%
       \put(\xpos,\ypos){\vector(0,-1)3}%
       \advance\ypos by-\halflength
       \advance\ypos by-\gap
       \put(\xpos,\ypos){\vector(0,-1){\halflength}}%
   \or \put(\xpos,\ypos){\line(0,-1)\halflength}%
       \advance\ypos by-\halflength
       \advance\ypos by-\gap
       \put(\xpos,\ypos){\evector(0,-1){\halflength}}%
   \fi
\else \arrowtype=-\arrowtype
   \ifcase\arrowtype
   \or \advance \ypos by-\arrowlength
       \put(\xpos,\ypos){\line(0,1){\halflength}}%
       \advance\ypos by\halflength
       \advance\ypos by\gap
       \put(\xpos,\ypos){\vector(0,1){\halflength}}%
   \or \advance \ypos by-\arrowlength
       \put(\xpos,\ypos){\line(0,1)\halflength}%
       \put(\xpos,\ypos){\vector(0,1)3}%
       \advance\ypos by\halflength
       \advance\ypos by\gap
       \put(\xpos,\ypos){\vector(0,1){\halflength}}%
   \or \advance \ypos by-\arrowlength
       \put(\xpos,\ypos){\line(0,1)\halflength}%
       \advance\ypos by\halflength
       \advance\ypos by\gap
       \put(\xpos,\ypos){\evector(0,1){\halflength}}%
   \fi
\fi
}

\def\putmorphism(#1)(#2,#3)[#4`#5`#6]#7#8#9{{%
\run #2
\rise #3
\ifnum\rise=0
  \puthmorphism(#1)[#4`#5`#6]{#7}{#8}#9%
\else\ifnum\run=0
  \putvmorphism(#1)[#4`#5`#6]{#7}{#8}#9%
\else
\setpos(#1)%
\arrowlength #7
\arrowtype #8
\ifnum\run=0
\else\ifnum\rise=0
\else
\ifnum\run>0
    \coefa=1
\else
   \coefa=-1
\fi
\ifnum\arrowtype>0
   \coefb=0
   \coefc=-1
\else
   \coefb=\coefa
   \coefc=1
   \arrowtype=-\arrowtype
\fi
\width=2
\multiply \width by\run
\divide \width by\rise
\ifnum \width<0  \width=-\width\fi
\advance\width by60
\if l#9 \width=-\width\fi
\putbox(\xpos,\ypos){#4}%            %node 1
{\multiply \coefa by\arrowlength%      %node 2
\advance\xpos by\coefa
\multiply \coefa by\rise
\divide \coefa by\run
\advance \ypos by\coefa
\putbox(\xpos,\ypos){#5} }%
{\multiply \coefa by\arrowlength%      %label
\divide \coefa by2
\advance \xpos by\coefa
\advance \xpos by\width
\multiply \coefa by\rise
\divide \coefa by\run
\advance \ypos by\coefa
\if l#9%
   \putrbox(\xpos,\ypos){#6}%
\else\if r#9%
   \putlbox(\xpos,\ypos){#6}%
\fi\fi }%
{\multiply \rise by-\coefc%             %arrow
\multiply \run by-\coefc
\multiply \coefb by\arrowlength
\advance \xpos by\coefb
\multiply \coefb by\rise
\divide \coefb by\run
\advance \ypos by\coefb
\multiply \coefc by70
\advance \ypos by\coefc
\multiply \coefc by\run
\divide \coefc by\rise
\advance \xpos by\coefc
\multiply \coefa by140
\multiply \coefa by\run
\divide \coefa by\rise
\advance \arrowlength by\coefa
\ifcase\arrowtype
\or \put(\xpos,\ypos){\vector(\run,\rise){\arrowlength}}%
\or \put(\xpos,\ypos){\mvector(\run,\rise){\arrowlength}}%
\or \put(\xpos,\ypos){\evector(\run,\rise){\arrowlength}}%
\fi}\fi\fi\fi\fi}}

\newcount\numbdashes \newcount\lengthdash \newcount\increment

\def\howmanydashes{% Actually returns both number and length
\numbdashes=\arrowlength \lengthdash=40
\divide\numbdashes by \lengthdash
\lengthdash=\arrowlength
\divide\lengthdash by \numbdashes
\increment=\lengthdash
\multiply\lengthdash by 3
\divide\lengthdash by 5
}

\def\putdashvector(#1)(#2,#3)#4#5{%
\ifnum#3=0 \putdashhvector(#1){#4}#5
\else
\ifnum#2=0
\putdashvvector(#1){#4}#5\fi\fi}

\def\putdashhvector(#1,#2)#3#4{{%
\arrowlength=#3 \howmanydashes
\multiput(#1,#2)(\increment,0){\numbdashes}%
{\vrule height .4pt width \lengthdash\unitlength}
\arrowtype=#4 \xpos=#1
\ifnum\arrowtype<0 \advance\arrowtype by 7 \fi
\ifcase\arrowtype
\or \advance\xpos by 10
    \put(\xpos,#2){\vector(-1,0){\lengthdash}}
    \advance\xpos by 40
    \put(\xpos,#2){\vector(-1,0){\lengthdash}}
\or \advance \xpos by 10
    \put(\xpos,#2){\vector(-1,0){\lengthdash}}
    \advance\xpos by  \arrowlength
    \advance\xpos by  -50
    \put(\xpos,#2){\vector(-1,0){\lengthdash}}
\or \advance\xpos by 10
    \put(\xpos,#2){\vector(-1,0){\lengthdash}}
\or \advance\xpos by \arrowlength
    \advance\xpos by -\lengthdash
    \put(\xpos,#2){\vector(1,0){\lengthdash}}
\or {\advance\xpos by 10
    \put(\xpos,#2){\vector(1,0){\lengthdash}}}
    \advance\xpos by \arrowlength
    \advance\xpos by -\lengthdash
    \put(\xpos,#2){\vector(1,0){\lengthdash}}
\or \advance\xpos by \arrowlength
    \advance\xpos by -\lengthdash
    \put(\xpos,#2){\vector(1,0){\lengthdash}}
    \advance\xpos by -40
    \put(\xpos,#2){\vector(1,0){\lengthdash}}
   \fi
}}

\def\putdashvvector(#1,#2)#3#4{{%
\arrowlength=#3 \howmanydashes
\ypos=#2 \advance\ypos by -\arrowlength
\multiput(#1,#2)(0,\increment){\numbdashes}%
    {\vrule width .4pt height \lengthdash\unitlength}
\arrowtype=#4 \ypos=#2
\ifnum\arrowtype<0 \advance\arrowtype by 7 \fi
\ifcase\arrowtype
\or \advance\ypos by \arrowlength \advance\ypos by -40
    \put(#1,\ypos){\vector(0,1){\lengthdash}}
    \advance\ypos by -40
    \put(#1,\ypos){\vector(0,1){\lengthdash}}
\or \advance\ypos by 10
    \put(#1,\ypos){\vector(0,1){\lengthdash}}
    \advance\ypos by \arrowlength \advance\ypos by -40
    \put(#1,\ypos){\vector(0,1){\lengthdash}}
\or \advance\ypos by \arrowlength \advance\ypos by -40
    \put(#1,\ypos){\vector(0,1){\lengthdash}}
\or \advance\ypos by 10
    \put(#1,\ypos){\vector(0,-1){\lengthdash}}
\or \advance\ypos by 10
    \put(#1,\ypos){\vector(0,-1){\lengthdash}}
    \advance\ypos by \arrowlength \advance\ypos by -40
    \put(#1,\ypos){\vector(0,-1){\lengthdash}}
\or \advance\ypos by 10
    \put(#1,\ypos){\vector(0,-1){\lengthdash}}
    \advance\ypos by 40
    \put(#1,\ypos){\vector(0,-1){\lengthdash}}
\fi
}}

\def\puthmorphism(#1,#2)[#3`#4`#5]#6#7#8{{%
\xpos #1
\ypos #2
\width #6
\arrowlength #6
\arrowtype=#7
\putbox(\xpos,\ypos){#3\vphantom{#4}}%
{\advance \xpos by\arrowlength
\putbox(\xpos,\ypos){\vphantom{#3}#4}}%
\horsize{\tempcounta}{#3}%
\horsize{\tempcountb}{#4}%
\divide \tempcounta by2
\divide \tempcountb by2
\advance \tempcounta by30
\advance \tempcountb by30
\advance \xpos by\tempcounta
\advance \arrowlength by-\tempcounta
\advance \arrowlength by-\tempcountb
\putvector(\xpos,\ypos)(1,0)\arrowlength\arrowtype
\divide \arrowlength by2
\advance \xpos by\arrowlength
\vertsize{\tempcounta}{#5}%
\divide\tempcounta by2
\advance \tempcounta by20
\if a#8 %
   \advance \ypos by\tempcounta
   \putbox(\xpos,\ypos){#5}%
\else
   \advance \ypos by-\tempcounta
   \putbox(\xpos,\ypos){#5}%
\fi}}

\def\putvmorphism(#1,#2)[#3`#4`#5]#6#7#8{{%
\xpos #1
\ypos #2
\arrowlength #6
\arrowtype #7
\settowidth{\xlen}{$#5$}%
\putbox(\xpos,\ypos){#3}%
{\advance \ypos by-\arrowlength
\putbox(\xpos,\ypos){#4}}%
{\advance\arrowlength by-140
\advance \ypos by-70
\ifdim\xlen>0pt
   \if m#8%
      \putsplitvector(\xpos,\ypos)\arrowlength\arrowtype
   \else
   \putvector(\xpos,\ypos)(0,-1)\arrowlength\arrowtype
   \fi
\else
   \putvector(\xpos,\ypos)(0,-1)\arrowlength\arrowtype
\fi}%
\ifdim\xlen>0pt
   \divide \arrowlength by2
   \advance\ypos by-\arrowlength
   \if l#8%
      \advance \xpos by-40
      \putrbox(\xpos,\ypos){#5}%
   \else\if r#8%
      \advance \xpos by40
      \putlbox(\xpos,\ypos){#5}%
   \else
      \putbox(\xpos,\ypos){#5}%
   \fi\fi
\fi
}}

\def\putsquarep<#1>(#2)[#3;#4`#5`#6`#7]{{%
\setsqparms[#1]%
\setpos(#2)%
\settokens`#3`%
\puthmorphism(\xpos,\ypos)[\tokenc`\tokend`{#7}]{\width}{\arrowtyped}b%
\advance\ypos by \height
\puthmorphism(\xpos,\ypos)[\tokena`\tokenb`{#4}]{\width}{\arrowtypea}a%
\putvmorphism(\xpos,\ypos)[``{#5}]{\height}{\arrowtypeb}l%
\advance\xpos by \width
\putvmorphism(\xpos,\ypos)[``{#6}]{\height}{\arrowtypec}r%
}}

\def\putsquare{\@ifnextchar <{\putsquarep}{\putsquarep%
   <\arrowtypea`\arrowtypeb`\arrowtypec`\arrowtyped;\width`\height>}}
\def\square{\@ifnextchar< {\squarep}{\squarep
   <\arrowtypea`\arrowtypeb`\arrowtypec`\arrowtyped;\width`\height>}}
\def\squarep<#1>[#2`#3`#4`#5;#6`#7`#8`#9]{{%       %     #2------>#3
\setsqparms[#1]%                                   %      |       |
\diagram%                                          %      |       |
\putsquarep<\arrowtypea`\arrowtypeb`\arrowtypec`%  %    #7|       |#8
\arrowtyped;\width`\height>%                       %      |       |
(0,0)[#2`#3`#4`{#5};#6`#7`#8`{#9}]%                %      |       |
\enddiagram%                                       %      v       v
}}                                                 %     #4------>#5
\def\putptrianglep<#1>(#2,#3)[#4`#5`#6;#7`#8`#9]{{%
\settriparms[#1]%
\xpos=#2 \ypos=#3
\advance\ypos by \height
\puthmorphism(\xpos,\ypos)[#4`#5`{#7}]{\height}{\arrowtypea}a%
\putvmorphism(\xpos,\ypos)[`#6`{#8}]{\height}{\arrowtypeb}l%
\advance\xpos by\height
\putmorphism(\xpos,\ypos)(-1,-1)[``{#9}]{\height}{\arrowtypec}r%
}}

\def\putptriangle{\@ifnextchar <{\putptrianglep}{\putptrianglep
   <\arrowtypea`\arrowtypeb`\arrowtypec;\height>}}
\def\ptriangle{\@ifnextchar <{\ptrianglep}{\ptrianglep
   <\arrowtypea`\arrowtypeb`\arrowtypec;\height>}}
\def\ptrianglep<#1>[#2`#3`#4;#5`#6`#7]{{%%    %      #2----->#3
\settriparms[#1]%                             %      |      /
\diagram%                                     %      |     /
\putptrianglep<\arrowtypea`\arrowtypeb`%      %    #6|    /#7
\arrowtypec;\height>%                         %      |   /
(0,0)[#2`#3`#4;#5`#6`{#7}]%                   %      |  /
\enddiagram%%                                 %      v v
}}                                            %      #4

\def\putqtrianglep<#1>(#2,#3)[#4`#5`#6;#7`#8`#9]{{%
\settriparms[#1]%
\xpos=#2 \ypos=#3
\advance\ypos by\height
\puthmorphism(\xpos,\ypos)[#4`#5`{#7}]{\height}{\arrowtypea}a%
\putmorphism(\xpos,\ypos)(1,-1)[``{#8}]{\height}{\arrowtypeb}l%
\advance\xpos by\height
\putvmorphism(\xpos,\ypos)[`#6`{#9}]{\height}{\arrowtypec}r%
}}

\def\putqtriangle{\@ifnextchar <{\putqtrianglep}{\putqtrianglep
   <\arrowtypea`\arrowtypeb`\arrowtypec;\height>}}
\def\qtriangle{\@ifnextchar <{\qtrianglep}{\qtrianglep
   <\arrowtypea`\arrowtypeb`\arrowtypec;\height>}}
\def\qtrianglep<#1>[#2`#3`#4;#5`#6`#7]{{%%    %        #2----->#3
\settriparms[#1]%                             %         \      |
\width=\height                                %          \     |
\diagram%                                     %         #6\    |#7
\putqtrianglep<\arrowtypea`\arrowtypeb`%      %            \   |
\arrowtypec;\height>%                         %             \  |
(0,0)[#2`#3`#4;#5`#6`{#7}]%                   %              v v
\enddiagram%%                                 %               #4
}}

\def\putdtrianglep<#1>(#2,#3)[#4`#5`#6;#7`#8`#9]{{%
\settriparms[#1]%
\xpos=#2 \ypos=#3
\puthmorphism(\xpos,\ypos)[#5`#6`{#9}]{\height}{\arrowtypec}b%
\advance\xpos by \height \advance\ypos by\height
\putmorphism(\xpos,\ypos)(-1,-1)[``{#7}]{\height}{\arrowtypea}l%
\putvmorphism(\xpos,\ypos)[#4``{#8}]{\height}{\arrowtypeb}r%
}}

\def\putdtriangle{\@ifnextchar <{\putdtrianglep}{\putdtrianglep
   <\arrowtypea`\arrowtypeb`\arrowtypec;\height>}}
\def\dtriangle{\@ifnextchar <{\dtrianglep}{\dtrianglep
   <\arrowtypea`\arrowtypeb`\arrowtypec;\height>}}
\def\dtrianglep<#1>[#2`#3`#4;#5`#6`#7]{{%%    %                  / |
\settriparms[#1]%                             %                 /  |
\width=\height                                %              #5/   |#6
\diagram%                                     %               /    |
\putdtrianglep<\arrowtypea`\arrowtypeb`%      %              /     |
\arrowtypec;\height>%                         %             v      v
(0,0)[#2`#3`#4;#5`#6`{#7}]%                   %            #3----->#4
\enddiagram%%                                 %                #7
}}

\def\putbtrianglep<#1>(#2,#3)[#4`#5`#6;#7`#8`#9]{{%
\settriparms[#1]%
\xpos=#2 \ypos=#3
\puthmorphism(\xpos,\ypos)[#5`#6`{#9}]{\height}{\arrowtypec}b%
\advance\ypos by\height
\putmorphism(\xpos,\ypos)(1,-1)[``{#8}]{\height}{\arrowtypeb}r%
\putvmorphism(\xpos,\ypos)[#4``{#7}]{\height}{\arrowtypea}l%
}}

\def\putbtriangle{\@ifnextchar <{\putbtrianglep}{\putbtrianglep
   <\arrowtypea`\arrowtypeb`\arrowtypec;\height>}}
\def\btriangle{\@ifnextchar <{\btrianglep}{\btrianglep
   <\arrowtypea`\arrowtypeb`\arrowtypec;\height>}}
\def\btrianglep<#1>[#2`#3`#4;#5`#6`#7]{{%%   %              | \
\settriparms[#1]%                            %              |  \
\width=\height                               %            #5|   \#6
\diagram%                                    %              |    \
\putbtrianglep<\arrowtypea`\arrowtypeb`%     %              |     \
\arrowtypec;\height>%                        %              v      v
(0,0)[#2`#3`#4;#5`#6`{#7}]%                  %              #3----->#4
\enddiagram%%                                %                 #7
}}

\def\putAtrianglep<#1>(#2,#3)[#4`#5`#6;#7`#8`#9]{{%
\settriparms[#1]%
\xpos=#2 \ypos=#3
{\multiply \height by2
\puthmorphism(\xpos,\ypos)[#5`#6`{#9}]{\height}{\arrowtypec}b}%
\advance\xpos by\height \advance\ypos by\height
\putmorphism(\xpos,\ypos)(-1,-1)[#4``{#7}]{\height}{\arrowtypea}l%
\putmorphism(\xpos,\ypos)(1,-1)[``{#8}]{\height}{\arrowtypeb}r%
}}

\def\putAtriangle{\@ifnextchar <{\putAtrianglep}{\putAtrianglep
   <\arrowtypea`\arrowtypeb`\arrowtypec;\height>}}
\def\Atriangle{\@ifnextchar <{\Atrianglep}{\Atrianglep
   <\arrowtypea`\arrowtypeb`\arrowtypec;\height>}}
\def\Atrianglep<#1>[#2`#3`#4;#5`#6`#7]{{%%         %         /   \
\settriparms[#1]%                                  %        /     \
\width=\height                                     %     #5/       \#6
\diagram%                                          %      /         \
\putAtrianglep<\arrowtypea`\arrowtypeb`%           %     /           \
\arrowtypec;\height>%                              %    v             v
(0,0)[#2`#3`#4;#5`#6`{#7}]%                        %   #3------------>#4
\enddiagram%%                                      %          #7
}}

\def\putAtrianglepairp<#1>(#2)[#3;#4`#5`#6`#7`#8]{{%
\settripairparms[#1]%
\setpos(#2)%
\settokens`#3`%
\puthmorphism(\xpos,\ypos)[\tokenb`\tokenc`{#7}]{\height}{\arrowtyped}b%
\advance\xpos by\height
\puthmorphism(\xpos,\ypos)[\phantom{\tokenc}`\tokend`{#8}]%
{\height}{\arrowtypee}b%
\advance\ypos by\height
\putmorphism(\xpos,\ypos)(-1,-1)[\tokena``{#4}]{\height}{\arrowtypea}l%
\putvmorphism(\xpos,\ypos)[``{#5}]{\height}{\arrowtypeb}m%
\putmorphism(\xpos,\ypos)(1,-1)[``{#6}]{\height}{\arrowtypec}r%
}}

\def\putAtrianglepair{\@ifnextchar <{\putAtrianglepairp}{\putAtrianglepairp%
   <\arrowtypea`\arrowtypeb`\arrowtypec`\arrowtyped`\arrowtypee;\height>}}
\def\Atrianglepair{\@ifnextchar <{\Atrianglepairp}{\Atrianglepairp%
   <\arrowtypea`\arrowtypeb`\arrowtypec`\arrowtyped`\arrowtypee;\height>}}

\def\Atrianglepairp<#1>[#2;#3`#4`#5`#6`#7]{{%           %  #2a
\settripairparms[#1]%                         %           / | \
\settokens`#2`%                               %          /  |  \
\width=\height                                %       #3/  #4   \#5
\diagram%                                     %        /    |    \
\putAtrianglepairp                            %       /     |     \
<\arrowtypea`\arrowtypeb`\arrowtypec`%        %      v      v      v
\arrowtyped`\arrowtypee;\height>%             %     #2b---->#2c---->#2d
(0,0)[{#2};#3`#4`#5`#6`{#7}]%                 %         #6     #7
\enddiagram%%
}}

\def\putVtrianglep<#1>(#2,#3)[#4`#5`#6;#7`#8`#9]{{%
\settriparms[#1]%
\xpos=#2 \ypos=#3
\advance\ypos by\height
{\multiply\height by2
\puthmorphism(\xpos,\ypos)[#4`#5`{#7}]{\height}{\arrowtypea}a}%
\putmorphism(\xpos,\ypos)(1,-1)[`#6`{#8}]{\height}{\arrowtypeb}l%
\advance\xpos by\height
\advance\xpos by\height
\putmorphism(\xpos,\ypos)(-1,-1)[``{#9}]{\height}{\arrowtypec}r%
}}

\def\putVtriangle{\@ifnextchar <{\putVtrianglep}{\putVtrianglep
   <\arrowtypea`\arrowtypeb`\arrowtypec;\height>}}
\def\Vtriangle{\@ifnextchar <{\Vtrianglep}{\Vtrianglep
   <\arrowtypea`\arrowtypeb`\arrowtypec;\height>}}
\def\Vtrianglep<#1>[#2`#3`#4;#5`#6`#7]{{%%     %        #2------------->#3
\settriparms[#1]%                              %         \             /
\width=\height                                 %          \           /
\diagram%                                      %         #6\         /#7
\putVtrianglep<\arrowtypea`\arrowtypeb`%       %            \       /
\arrowtypec;\height>%                          %             \     /
(0,0)[#2`#3`#4;#5`#6`{#7}]%                    %              v   v
\enddiagram%%                                  %               #4
}}

\def\putVtrianglepairp<#1>(#2)[#3;#4`#5`#6`#7`#8]{{
\settripairparms[#1]%
\setpos(#2)%
\settokens`#3`%
\advance\ypos by\height
\putmorphism(\xpos,\ypos)(1,-1)[`\tokend`{#6}]{\height}{\arrowtypec}l%
\puthmorphism(\xpos,\ypos)[\tokena`\tokenb`{#4}]{\height}{\arrowtypea}a%
\advance\xpos by\height
\puthmorphism(\xpos,\ypos)[\phantom{\tokenb}`\tokenc`{#5}]%
{\height}{\arrowtypeb}a%
\putvmorphism(\xpos,\ypos)[``{#7}]{\height}{\arrowtyped}m%
\advance\xpos by\height
\putmorphism(\xpos,\ypos)(-1,-1)[``{#8}]{\height}{\arrowtypee}r%
}}

\def\putVtrianglepair{\@ifnextchar <{\putVtrianglepairp}{\putVtrianglepairp%
    <\arrowtypea`\arrowtypeb`\arrowtypec`\arrowtyped`\arrowtypee;\height>}}
\def\Vtrianglepair{\@ifnextchar <{\Vtrianglepairp}{\Vtrianglepairp%
    <\arrowtypea`\arrowtypeb`\arrowtypec`\arrowtyped`\arrowtypee;\height>}}
\def\Vtrianglepairp<#1>[#2;#3`#4`#5`#6`#7]{{%  %  #2a---->#2b---->#2c
\settripairparms[#1]%                          %   \      |      /
\settokens`#2`%                                %    \     |     /
\diagram%                                      %   #5\   #6    /#7
\putVtrianglepairp                             %      \   |   /
<\arrowtypea`\arrowtypeb`\arrowtypec`%         %       \  |  /
\arrowtyped`\arrowtypee;\height>%              %        v v v
(0,0)[{#2};#3`#4`#5`#6`{#7}]%                  %         #2d
\enddiagram%%
}}

\def\putCtrianglep<#1>(#2,#3)[#4`#5`#6;#7`#8`#9]{{%
\settriparms[#1]%
\xpos=#2 \ypos=#3
\advance\ypos by\height
\putmorphism(\xpos,\ypos)(1,-1)[``{#9}]{\height}{\arrowtypec}l%
\advance\xpos by\height
\advance\ypos by\height
\putmorphism(\xpos,\ypos)(-1,-1)[#4`#5`{#7}]{\height}{\arrowtypea}l%
{\multiply\height by 2
\putvmorphism(\xpos,\ypos)[`#6`{#8}]{\height}{\arrowtypeb}r}%
}}

\def\putCtriangle{\@ifnextchar <{\putCtrianglep}{\putCtrianglep
    <\arrowtypea`\arrowtypeb`\arrowtypec;\height>}}
\def\Ctriangle{\@ifnextchar <{\Ctrianglep}{\Ctrianglep
    <\arrowtypea`\arrowtypeb`\arrowtypec;\height>}}
\def\Ctrianglep<#1>[#2`#3`#4;#5`#6`#7]{{%%   %                / |
\settriparms[#1]%                            %             #5/  |
\width=\height                               %              /   |
\diagram%                                    %             v    |
\putCtrianglep<\arrowtypea`\arrowtypeb`%     %           #3     |#6
\arrowtypec;\height>%                        %             \    |
(0,0)[#2`#3`#4;#5`#6`{#7}]%                  %            #7\   |
\enddiagram%%                                %               \  |
}}                                           %                v v
\def\putDtrianglep<#1>(#2,#3)[#4`#5`#6;#7`#8`#9]{{%
\settriparms[#1]%
\xpos=#2 \ypos=#3
\advance\xpos by\height \advance\ypos by\height
\putmorphism(\xpos,\ypos)(-1,-1)[``{#9}]{\height}{\arrowtypec}r%
\advance\xpos by-\height \advance\ypos by\height
\putmorphism(\xpos,\ypos)(1,-1)[`#5`{#8}]{\height}{\arrowtypeb}r%
{\multiply\height by 2
\putvmorphism(\xpos,\ypos)[#4`#6`{#7}]{\height}{\arrowtypea}l}%
}}

\def\putDtriangle{\@ifnextchar <{\putDtrianglep}{\putDtrianglep
    <\arrowtypea`\arrowtypeb`\arrowtypec;\height>}}
\def\Dtriangle{\@ifnextchar <{\Dtrianglep}{\Dtrianglep
   <\arrowtypea`\arrowtypeb`\arrowtypec;\height>}}
\def\Dtrianglep<#1>[#2`#3`#4;#5`#6`#7]{{%%  %          | \
\settriparms[#1]%                           %          |  \#6
\width=\height                              %          |   \
\diagram%                                   %          |    v
\putDtrianglep<\arrowtypea`\arrowtypeb`%    %        #5|    #3
\arrowtypec;\height>%                       %          |    /
(0,0)[#2`#3`#4;#5`#6`{#7}]%                 %          |   /#7
\enddiagram%%                               %          |  /
}}                                          %          v v
\def\setrecparms[#1`#2]{\width=#1 \height=#2}%
\def\recursep<#1`#2>[#3;#4`#5`#6`#7`#8]{{\m@th
\width=#1 \height=#2
\settokens`#3`
\settowidth{\tempdimen}{$\tokena$}
\ifdim\tempdimen=0pt
  \savebox{\tempboxa}{\hbox{$\tokenb$}}%
  \savebox{\tempboxb}{\hbox{$\tokend$}}%
  \savebox{\tempboxc}{\hbox{$#6$}}%
\else
  \savebox{\tempboxa}{\hbox{$\hbox{$\tokena$}\times\hbox{$\tokenb$}$}}%
  \savebox{\tempboxb}{\hbox{$\hbox{$\tokena$}\times\hbox{$\tokend$}$}}%
  \savebox{\tempboxc}{\hbox{$\hbox{$\tokena$}\times\hbox{$#6$}$}}%
\fi
\ypos=\height
\divide\ypos by 2
\xpos=\ypos
\advance\xpos by \width
\bfig
\putCtrianglep<-1`1`1;\ypos>(0,0)[`\tokenc`;#5`#6`{#7}]%
\puthmorphism(\ypos,0)[\tokend`\usebox{\tempboxb}`{#8}]{\width}{-1}b%
\puthmorphism(\ypos,\height)[\tokenb`\usebox{\tempboxa}`{#4}]{\width}{-1}a%
\advance\ypos by \width
\putvmorphism(\ypos,\height)[``\usebox{\tempboxc}]{\height}1r%
\efig
}}

\def\recurse{\@ifnextchar <{\recursep}{\recursep<\width`\height>}}

\def\puttwohmorphisms(#1,#2)[#3`#4;#5`#6]#7#8#9{{%
\puthmorphism(#1,#2)[#3`#4`]{#7}0a
\ypos=#2
\advance\ypos by 20
\puthmorphism(#1,\ypos)[\phantom{#3}`\phantom{#4}`#5]{#7}{#8}a
\advance\ypos by -40
\puthmorphism(#1,\ypos)[\phantom{#3}`\phantom{#4}`#6]{#7}{#9}b
}}

\def\puttwovmorphisms(#1,#2)[#3`#4;#5`#6]#7#8#9{{%
\putvmorphism(#1,#2)[#3`#4`]{#7}0a
\xpos=#1
\advance\xpos by -20
\putvmorphism(\xpos,#2)[\phantom{#3}`\phantom{#4}`#5]{#7}{#8}l
\advance\xpos by 40
\putvmorphism(\xpos,#2)[\phantom{#3}`\phantom{#4}`#6]{#7}{#9}r
}}

\def\puthcoequalizer(#1)[#2`#3`#4;#5`#6`#7]#8#9{{%
\setpos(#1)%
\puttwohmorphisms(\xpos,\ypos)[#2`#3;#5`#6]{#8}11%
\advance\xpos by #8
\puthmorphism(\xpos,\ypos)[\phantom{#3}`#4`#7]{#8}1{#9}
}}

\def\putvcoequalizer(#1)[#2`#3`#4;#5`#6`#7]#8#9{{%
\setpos(#1)%
\puttwovmorphisms(\xpos,\ypos)[#2`#3;#5`#6]{#8}11%
\advance\ypos by -#8
\putvmorphism(\xpos,\ypos)[\phantom{#3}`#4`#7]{#8}1{#9}
}}

\def\putthreehmorphisms(#1)[#2`#3;#4`#5`#6]#7(#8)#9{{%
\setpos(#1) \settypes(#8)
\if a#9 %
     \vertsize{\tempcounta}{#5}%
     \vertsize{\tempcountb}{#6}%
     \ifnum \tempcounta<\tempcountb \tempcounta=\tempcountb \fi
\else
     \vertsize{\tempcounta}{#4}%
     \vertsize{\tempcountb}{#5}%
     \ifnum \tempcounta<\tempcountb \tempcounta=\tempcountb \fi
\fi
\advance \tempcounta by 60
\puthmorphism(\xpos,\ypos)[#2`#3`#5]{#7}{\arrowtypeb}{#9}
\advance\ypos by \tempcounta
\puthmorphism(\xpos,\ypos)[\phantom{#2}`\phantom{#3}`#4]{#7}{\arrowtypea}{#9}
\advance\ypos by -\tempcounta \advance\ypos by -\tempcounta
\puthmorphism(\xpos,\ypos)[\phantom{#2}`\phantom{#3}`#6]{#7}{\arrowtypec}{#9}
}}

\def\setarrowtoks[#1`#2`#3`#4`#5`#6]{%
\def\toka{#1}
\def\tokb{#2}
\def\tokc{#3}
\def\tokd{#4}
\def\toke{#5}
\def\tokf{#6}
}
\def\hex{\@ifnextchar <{\hexp}{\hexp<1000`400>}}
\def\hexp<#1`#2>[#3`#4`#5`#6`#7`#8;#9]{%
\setarrowtoks[#9]
\yext=#2 \advance \yext by #2
\xext=#1 \advance\xext by \yext
\bfig
\putCtriangle<-1`0`1;#2>(0,0)[`#5`;\tokb``\tokd]
\xext=#1 \yext=#2 \advance \yext by #2
\putsquare<1`0`0`1;\xext`\yext>(#2,0)[#3`#4`#7`#8;\toka```\tokf]
\advance \xext by #2
\putDtriangle<0`1`-1;#2>(\xext,0)[`#6`;`\tokc`\toke]
\efig
}
\makeatother
\newtheorem{lemma}{Lemma}[section]

\newtheorem{theorem}[lemma]{Theorem}

\newcommand{\EProof}{\hfill \blacksquare}

\def\ome {\omega}

\def\XI{\xi}
\def\XII{\eta}
\def\CHI{\chi}

\def\lto{\longrightarrow}

\def\iso{\backsimeq}

\def\Z{{\mathbb Z}}
\def\C{{\mathbb C}}
\def\lgn {\sigma}  %  Legendre character
\def\Fp{\mathbb{F}_{p}}

\def\Lm{\Lambda^*}

\def\GL{\mathrm{GL}}

\def\GZ{\mathrm{SL}_2(\Z)}
\def\G{\mathrm{\Gamma}}     %{\mathrm G}
\def\GG{\mathrm{SL}_2(\Fp)}

\def\AGG{\mathbb{S} \mathbb{L}_2} % algebraic group SL2
\def\AT{\mathbb{T}}      % algebraic standard torus

\def\CA{{\mathrm T}_A}
\def\ACA{\mathbb{T}_A}
\def\Gf{\mathrm{\Gamma}_{p}}

\def\Sp{\mathrm{Sp}} % appendix A, group of linear symplectomorphisms of the f.d v.s
\def\ASp{\mathbb{S}\mathrm{p}} % algebraic group of linear symplectomorphisms
\def\S0{S}

\def\F{C^{\infty}({\bf T})}
\def\T{{\bf T}}
\def\Td{\T^\vee} % the lattice of characters on T

\def\A{\cal A}
\def\Ad{ {\cal A} _\hbar}

\def\h{ \hbar }

\def\Hh{{\cal H} _\h}
\def\V{\mathrm{V}} % symplectic vector space
\def\AV {\mathbb{V}} % algebraic variety corresponding to V
\def\W{{\mathrm W}}

\def\rhoh{\rho _{_\h}}
\def\pih{\pi i {\h}}
\def\Pih{\pi_{_\hbar}}

\def\Y{Y_{0}}
\def\YY{Y}
\def\AYY{\mathbb{Y}}

\def \bA{\mathbb{A}}
\def\P1{\mathbb{P}^1}

\def\AX {\mathbb{X}} % a general variety

\def\i_XI{i_{_{\XI}}}
\def\iXII{i_{_{\XII}}}
\def\p_XI{p_{_{\XI}}}

\def\SF{\mathcal{F}}
\def\SG{\mathcal{G}}
\def\SL{\mathscr L}

\def\Skummer {\SL_{\chi}} % Kummer sheaf

\def\Tr{\mathrm{Tr}}
\def\Av{\mathrm{\bf Av}}

\def\End{\mathrm{End}}
\def\Hom{\mathrm{Hom}}

\def\Db{{\cal D}^{b}_{\mathrm{c},\mathrm{w}}}
\def\Fr{\mathrm{Fr}}

\def\coH{{\mathrm H}}

\def\r{\;}

\def\half{\begin{smallmatrix} \frac{1}{2} \end{smallmatrix}}

\begin{document}

\title{\bf \small PROOF OF THE KURLBERG-RUDNICK RATE CONJECTURE}
\author{\small\it SHAMGAR GUREVICH AND RONNY HADANI}

\date{}

\maketitle

{\narrower{\narrower{\narrower{\bigskip\noindent{Abstract.}\enspace
In this paper we present a proof of the {\it Hecke quantum unique
ergodicity conjecture} for the Berry-Hannay model, a model of
quantum mechanics on a two dimensional torus. This conjecture was
stated in Z. Rudnick's lectures at MSRI, Berkeley, 1999 and ECM,
Barcelona, 2000.}}}}
\spc{1}
{\narrower{\narrower{\narrower{\bigskip\bigskip\bigskip\noindent
{R\'esum\'e.}\enspace Nous proposons une d\'emonstration de la
conjecture d'unique ergodicit\'e quantique d'Hecke pour le
mod\`{e}le de Berry-Hannay, un mod\`{e}le  de m\'{e}canique
quantique sur un tore de dimension deux. Cette conjecture  a
\'{e}t\'{e} propos\'{e}e par Z. Rudnick \'{a} MSRI, Berkeley, 1999
\`{a} l'ECM, Barcelona, 2000.}}}}
\numberwithin{equation}{subsection}

\setcounter{section}{-1}

\spc{1}
\section{Introduction}
\spc{0.2}
\textbf{Hannay-Berry model.} In 1980 the physicists J. Hannay and
Sir M.V. Berry \cite{HB} explore a model for quantum mechanics on
the two dimensional symplectic torus $(\T,\ome)$.
\\
\\
\textbf{Quantum chaos.} Consider the ergodic discrete dynamical
system on the torus, which is generated by an hyperbolic
automorphism $A \in \GZ$. Quantizing the system, we replace: the
classical phase space $(\T,\ome)$  by a Hilbert space $\Hh$,
classical observables, i.e., functions $f \in \F$, by operators
$\Pih(f) \in \End(\Hh)$ and classical symmetries by a unitary
representation $\rhoh : \GZ \lto \mathrm{U}(\Hh)$. A fundamental
meta-question in the area of quantum chaos is to \textit{describe}
the spectral properties of the quantum system $\rhoh(A)$, at least
in the semi-classical limit as $\h \rightarrow 0$.

\textbf{The rate conjecture.} In \cite{KR1} Kurlberg and Rudnick
proved that eigenvectors that satisfy certain additional
symmetries of $\rhoh(A)$ are semi-classically equidistributed with
respect to the Haar measure on $\T$. In this paper we prove (see
Theorem \ref{QHUE}) the Kurlberg-Rudnick conjecture \cite{R1, R2}
on the rate of convergence of the relevant distribution to the
Haar measure.
\spc{0.2}
\subsection*{Acknowledgments}
\spc{0.2} We thank our Ph.D. adviser J. Bernstein for his interest
and guidance in this project. We thank P. Kurlberg and Z. Rudnick
who discussed with us their papers and explained their results. We
would like to thank David Kazhdan for sharing his thoughts about
the possible existence of canonical Hilbert spaces. Finally, we
would like to thank P. Deligne for letting us publish his ideas
about the geometrization of the Weil representation which appeared
in a letter \cite{D1} he wrote to David Kazhdan in 1982.
\spc{0.2}
\section{Classical Torus} \label{classicaltorus}
\spc{0.2} Let $(\T,\ome)$ be the two dimensional symplectic torus.
Together with its linear symplectomorphisms $\G \iso \GZ$ it
serves as a simple model of classical mechanics (a compact version
of the phase space of the harmonic oscillator). More precisely,
let $\T= \W/ \Lambda$ where $\W$ is a two dimensional real vector
space and $\Lambda$ is a rank two unimodular lattice in $\W$. We
denote by $\Lm \subseteq \W^*$ the dual lattice, i.e., $\Lm = \{
\xi \in \W^* | \r\r \xi (\Lambda) \subset \Z \} $. The lattice
$\Lm$ is identified with the lattice of characters of $\T$ by the
map $\xi \in \Lm \longmapsto e^{2 \pi i <\xi, \cdot>} \in \r \Td,$
where $\Td := \Hom(\T,\C^*)$.
\\
\\
\textbf{Classical mechanical system.} We consider a very simple
discrete mechanical system. An hyperbolic element $A \in \G$,
i.e., $|\Tr(A)| > 2$, generates an ergodic discrete dynamical
system on $\T$.
\spc{0.2}
\section{Quantization of the Torus} \label{quantization}
\spc{0.2}
\textbf{The Weyl quantization model.} The Weyl quantization model
works as follows. Let $\Ad$ be a one parameter deformation of the
algebra $\A$ of trigonometric polynomials on the torus. This
algebra is known in the literature as the Rieffel torus \cite{Ri}.
The algebra $\Ad $ is constructed by taking the free algebra over
$\C$ generated by the symbols $\{s(\xi) \r | \r \xi \in \Lm \}$
and quotient out by the relation $ s(\xi + \eta) = e^{\pih
\ome(\xi,\eta)}s(\xi)s(\eta)$. Here $\ome$ is the form on $\W^*$
induced by the original form $\ome$ on $\W$. The algebra $\Ad$
contains as a standard basis the lattice $\Lm$. Therefore, one can
identify the algebras $\Ad \simeq \A$ as vector spaces. Hence,
every function $f \in \A$ can be viewed as an element of $\Ad$.
For a fixed $\hbar$ a representation $\Pih:\Ad \lto \End(\Hh)$
serves as a quantization protocol.
\\
\\
\textbf{Equivariant Weyl quantization of the torus.} The group
$\G$ acts on the lattice $\Lm$, therefore it acts on $\Ad$. For an
element $B \in \G$, we denote by $ f \longmapsto f^B$ the action
of $B$ on an element $f \in \Ad$. Let $\Gf \iso \GG$ denotes the
quotient group of $\G$ modulo $p$.

\begin{theorem}[Canonical equivariant quantization] Let $\hbar =
\frac{1}{p}$, where $p$ is an odd prime. There exists a
\textit{unique} $($up to isomorphism$)$ pair of representations
$\Pih : \Ad \lto \End(\Hh)$ and $\rhoh : \G  \lto  \GL(\Hh)$
satisfying the compatibility condition $($\textit{Egorov
identity}$)$ $\rhoh(B) \Pih(f) \rhoh(B)^{-1} = \Pih(f^B)$, where
$\Pih$ is an irreducible representation and $\rhoh$ is a
representation of $\G$ that factors through the quotient group
$\Gf$.
\end{theorem}
\textbf{Quantum mechanical system.} Let $(\Pih,\rhoh,\Hh)$ be the
canonical equivariant quantization. Let $A$ be our fixed
hyperbolic element, considered as an element of $\Gf$ . The
element $A$ generates a quantum dynamical system. For every (pure)
quantum state $v \in S(\Hh) = \{ v \in \Hh :\|v\|=1\},\; v
\longmapsto v^A := \rhoh(A)v$.
\spc{0.2}
\section{Hecke Quantum Unique Ergodicity} \label{QHUE}
\spc{0.2} Denote by $\CA$ the centralizer of $A$ in $\Gf \iso
\GG$. We call $\CA$ the {\it Hecke torus} (cf. \cite{KR1}). The
precise statement of the \textbf{Kurlberg-Rudnick conjecture} (cf.
\cite{DGI} and \cite{R1, R2}) is given in the following theorem:
\begin{theorem}[Hecke Quantum Unique Ergodicity]\label{GH3} Let $\hbar =
\frac{1}{p}, \r p$ an odd prime. For every $f \in \Ad$ and $v \in
S(\Hh)$, we have:
\begin{equation}\label{qheckerg}
\left| \Av_{_{\CA}} ( <v| \Pih(f) v> ) - \int_{\T}f \ome \right|
\leq \frac{C_{f}}{\sqrt{p}},
\end{equation}
where $\Av_{_{\CA}}(<v|\Pih(f)v>) := \sum\limits_{B \in \CA}
<v|\Pih(f^B)v>$ is the average with respect to the group $\CA$ and
$C_{f}$ is an explicit constant depending only on $f$.
\end{theorem}
\spc{0.2}
\section{Proof of the Hecke Quantum Unique Ergodicity Conjecture}\label{Proof}
\spc{0.2} It is enough to prove the conjecture for the case when
$f$ is a non-trivial character $\xi \in \Lambda ^*$ and $v$ is an
Hecke eigenvector with eigencharacter $\chi:\CA \lto \C^*$. In
this case Theorem \ref{GH3} can be restated in the form:
\begin{theorem}[Hecke Quantum Unique Ergodicity (Restated)]\label{GH4}
Let $\hbar = \frac{1}{p}$, where $p$ is an odd prime. For every
$\xi \in \Lm$ and every character $\chi:\CA \lto \C^*$ the
following holds:
\begin{equation*}
\left| \sum_{B \in \CA} \Tr( \rhoh(B) \Pih(\xi)) \chi(B) \right|
\leq 2 \sqrt{p}.
\end{equation*}
\end{theorem}
\textbf{The trace function.} Denote by $F$ the function $F : \G
\times \Lm \lto \C$ defined by $F(B,\xi) = \Tr(\rho(B)
\Pih(\xi))$. We denote by $\V := \Lm / p \Lm$ the quotient vector
space, i.e., $\V \simeq \Fp^2$. The symplectic form $\ome$
specializes to give a symplectic form on $\V$. The group $\Gf$ is
the group of linear symplectomorphisms of $\V$, i.e., $\Gf =
\Sp(\V,\ome)$. Set $\Y := \G \times \Lm$ and $\YY := \Gf \times
\V$. We have a natural quotient map $\Y \lto \YY$.
\begin{lemma}\label{factorization}
The function $F : \Y \lto \C$ factors through the quotient $\YY$.
\end{lemma}
From now on $\YY$ will be considered as the default domain of the
function $F$. The function $F:\YY \lto \C$ is invariant with
respect to the action of $\Gf$ on $\YY$ given by the following
formula:
\begin{equation}\label{actionset}
  \begin{CD}
   \Gf \times \YY    @>\alpha>> \YY,\\
   (S,(B, \xi))      @>>>  (SBS^{-1} , S \xi).
   \end{CD}
\end{equation}
\textbf{Geometrization (Sheafification).} Next, we will phrase a
geometric statement that will imply Theorem \ref{GH4}. Moving into
the geometric setting, we replace the set $\YY$ by an algebraic
variety and the functions $F$ and $\CHI$ by
sheaf theoretic objects, also of a geometric flavor.\\\\
\textbf{Step 1.}  The set $\YY$ is the set of rational points of
an  algebraic variety $\AYY$ defined over $\Fp$. To be more
precise, $\AYY \simeq \ASp \times \AV$. The variety $\AYY$ is
equipped with an endomorphism $\Fr:\AYY \lto \AYY$ called
Frobenius. The set $\YY$ is identified with the set of fixed
points of Frobenius $\YY = \AYY^{\Fr} = \{ y \in \AYY : \Fr(y) = y
\}.$ Finally, we denote by $\alpha$ the algebraic action of $\ASp$
on the variety $\AYY$ (cf. (\ref{actionset})).
\\
\\
%
%
%
%
%
% The theorem about the sheaf \Sf
\textbf{Step 2.} The following theorem proposes an appropriate
sheaf theoretic object standing in place of the function $F : \YY
\lto \C$. Denote by $\Db (\AYY)$  the bounded derived category of
constructible $\ell$-adic Weil sheaves on $\AYY$.
\begin{theorem}[Geometrization Theorem]\label{deligne}
There exists an object $\SF \in \Db(\AYY) $
satisfying the following properties:
\end{theorem}

\begin{enumerate}
\item \label{prop_del1}$($Function$)$ It is associated, via the \textit{sheaf-to-function correspondence}, to the
function
\\
$F : \YY \lto \C$, i.e., $f^{\SF} = F$.
\item \label{prop_del2} $($Weight$)$ It is of weight $w(\SF) \leq
0.$
\item \label{prop_del3} $($Equivariance$)$  For every element $S \in
  \ASp$ there exists an isomorphism $\alpha_S ^* \SF \simeq
  \SF.$
\item \label{prop_del4} $($Formula$)$ On introducing coordinates $\AV
  \simeq \bA^2$ we identify $\ASp \simeq \AGG$. Then there exists an
  isomorphism $\SF_{|_{\AT \times \AV}} \simeq \SL_{\psi(\half
\lambda\mu\frac{a+1}{a-1})} \otimes \SL_{\lgn(a)}$.\footnote{By
this we mean that $\SF_{|_{\AT \times \AV}}$ is isomorphic to the
extension of the sheaf defined by the formula in the right-hand
side.}
\\
Here $\AT := \{ \left(
\begin{smallmatrix} a & 0 \\ 0 & a^{-1}
\end{smallmatrix} \right) \}$ stands for the standard torus,
$(\lambda,\mu)$ are the coordinates on $\AV$ and $\SL_\psi,\;
\SL_\lgn$ the Artin-Schreier and Kummer sheaves.
\end{enumerate}
%
%The explanation of the Geometrization Theorem
%
%
\textbf{Geometric statement.} Fix an element $\xi \in \Lm$ with
$\xi \neq 0$. We denote by $\i_XI$ the inclusion map $\i_XI : \CA
\times \xi \lto \YY $.
%
% diagrammatic form of the HQUE Theorem
%
Going back to Theorem \ref{GH4} and putting its content in a
functorial notation, we write the following inequality:
\begin{equation*}
  \left| pr_! ( \i_XI ^* (F) \cdot \CHI) \right| \leq 2\sqrt {p}.
\end{equation*}
In words, taking the function $F : \YY \lto \C$ and restricting
$F$ to $\CA \times \xi $ and get $\i_XI^ * (F)$. Multiply $\i_XI^
* F $ by the character $\CHI$ to get $\i_XI^ * (F) \cdot \CHI$.
Integrate $\i_XI^ * (F)  \cdot \CHI$ to the point, this means to
sum up all its values, and get a scalar $a_{\CHI} := pr_! ( \i_XI^
* (F)  \cdot \CHI)$. Here $pr$ stands for
  the projection $pr : \CA \times \xi \lto pt$. Then Theorem \ref{GH4} asserts that the scalar $a_{\CHI}$ is of
  an absolute value less than $2\sqrt{p}$.\\\\
%
% The geometric form of HQUE Theorem
%
Repeat the same steps in the geometric setting. We denote again by
$\i_XI$ the closed imbedding  $\i_XI: \ACA \times \xi \lto \AYY $.
Take the sheaf $\SF$ on $\AYY$ and apply the following sequence of
operations. Pull-back $\SF$ to the closed subvariety $\ACA \times
\xi$ and get the sheaf $\i_XI^ * (\SF)$. Take the tensor product
of $\i_XI^ * (\SF)$ with the Kummer sheaf ${\SL}_{\CHI}$ and get
$\i_XI^ * (\SF) \otimes {\SL}_\CHI$. Integrate $\i_XI^ * (\SF)
\otimes {\SL}_\CHI$ to the point and get the sheaf $pr_! (\i_XI^ *
(\SF) \otimes {\SL}_\CHI)$ on the point.
\\
\\
Recall $w(\SF) \leq 0$. Knowing that the Kummer sheaf has weight
$w(\SL_{\CHI}) \leq 0$ we deduce that $w(\i_XI^*(\SF) \otimes
\SL_{\CHI}) \leq 0.$
%
% The affect of integration on Weight (Deligne Theorem)
%
%
\begin{theorem}[Deligne, Weil II \cite{D2}]\label{deligne2}
Let $\pi :\AX_1 \lto \AX_2$ be a morphism of algebraic varieties.
Let ${\cal L} \in \Db(\AX_1)$ be a sheaf of weight $w({\cal L})
\leq w$ then $w(\pi_! ({\cal L})) \leq w$.
\end{theorem}
Using Theorem \ref{deligne2} we get $w( pr_! (\i_XI^*(\SF) \otimes
\SL_{\CHI})) \leq 0.$
\\
\\
Now, consider the sheaf $\SG := pr_! (\i_XI^*(\SF) \otimes
\SL_{\CHI})$. It is an object in $\Db(pt)$. The sheaf $\SG$ is
associated by {\it Grothendieck's Sheaf-To-Function
correspondence} to the scalar $a_{\CHI}$:
\begin{equation}\label{geulerchar}
  a_{\CHI} = \sum_{i \in \Z} (-1)^i \Tr(\Fr \big{|}_{\coH^i(\SG)}).
\end{equation}
Finally, we can give the geometric statement  about $\SG$,  which will imply Theorem \ref{GH4}.
\begin{lemma}[Vanishing Lemma]\label{vanishing}
Let $\SG = pr_! (\i_XI^*(\SF) \otimes \SL_{\CHI})$. All
cohomologies $\coH^{i}(\SG)$ vanish except for $i=1$. Moreover,
$\coH^1(\SG)$ is a two dimensional vector space.
\end{lemma}
Theorem \ref{GH4}  now follows easily. By Lemma \ref{vanishing}
only the first cohomology $\coH^1(\SG)$ does not vanish and it is
two dimensional. Having that $w(\SG) \leq 0$ implies that the
eigenvalues of Frobenius acting on $\coH^1(\SG)$ are of absolute
value $ \leq \sqrt{p}$. Hence, using formula (\ref{geulerchar}) we
get $| a_{\CHI} | \leq 2 \sqrt{p}.$
\\
\\
\textbf{Proof of the Vanishing Lemma.} \textbf{Step 1.} All tori
in $\ASp$ are conjugated. On introducing coordinates, i.e., $\AV
\simeq \bA^2$, we make the identification $\ASp \simeq \AGG$.  In
these terms there exists an element $\S0 \in \AGG$ conjugating the
{\it Hecke} torus $\ACA \subset \AGG$ with the standard torus
$\AT = \left \{ ( \begin{smallmatrix} a & 0 \\
0 & a^{-1} \end{smallmatrix} \right ) \} \subset \AGG $, namely
$\S0 \ACA \S0^{-1} = \AT$.
\\
\\
%
%
% The reduction to the sheaf $\SG_{st}$.
%
\textbf{Step 2.} Using the equivariance property of the sheaf
$\SF$ (see Theorem \ref{deligne}, property  \ref{prop_del3}) we
see that it is \textit{sufficient} to prove the Vanishing Lemma
for the sheaf $\SG_{st} := pr_! (\iXII ^* \SF \otimes
{\alpha_{_\S0}}_! \Skummer )$, where $\XII = \S0 \cdot \XI$ and
$\alpha_{_{\S0}}$ is the restriction of the action $\alpha$ to the
element $\S0$.
\\
\\
\textbf{Step 3.} The Vanishing Lemma holds for the sheaf
$\SG_{st}$. We write $ \XII = (\lambda,\mu)$. By Theorem
\ref{deligne} Property \ref{prop_del4} we have $ \iXII^* \SF
\simeq \SL_{\psi( \half \lambda\mu \frac{a+1}{a-1})} \otimes
\SL_{\lgn(a)} $, where $a$ is the coordinate of the standard torus
$\AT$ and $\lambda \cdot \mu \neq 0$\footnote{This is a direct
  consequence of the fact that $A \in \GZ$ is an hyperbolic element and
  does not have eigenvectors in $\Lm$.}. The sheaf $ {\alpha_{_{\S0}}}_!  \Skummer $
  is a character sheaf on the torus $\AT$. A direct computation proves the Vanishing Lemma.$\EProof$
%
%
%%%%%%%%%%%%%%%%%%%%%%%%%%%%%%%%%%%%%%%

\end{document}